\documentclass[preprint,12pt]{aastex}

\newcommand{\be}{\begin{equation}}
\newcommand{\ee}{\end{equation}}

\newcommand{\bea}{\begin{eqnarray}}
\newcommand{\eea}{\end{eqnarray}}

\newcommand{\alphav}{\alpha}

\begin{document}
\title{Global Disk Oscillation Modes in Cataclysmic Variables and Other Newtonian Accretors}

\author{Manuel Ortega-Rodr\'{\i}guez\altaffilmark{1}}
\affil{Escuela de F\'{\i}sica \& Centro de Investigaciones
Geof\'{\i}sicas,  \\ Universidad de Costa Rica,  
San Jos\'e, Costa Rica; \\
and 
Gravity Probe B,
W. W. Hansen Experimental Physics Lab, \\
Stanford University, Stanford, CA 94305-4085
}
\author{Robert V. Wagoner\altaffilmark{2}}
\affil{Dept.~of Physics \& Kavli Institute for Particle Astrophysics and Cosmology,\\ 
Stanford University, Stanford, CA 94305--4060}
\altaffiltext{1}{mortega@cariari.ucr.ac.cr}
\altaffiltext{2}{wagoner@stanford.edu}

\begin{abstract}
Diskoseismology, the theoretical study of small adiabatic hydrodynamical global perturbations of geometrically thin, optically thick accretion disks around black holes (and other compact objects), is a potentially powerful probe of the gravitational field. For instance, the frequencies of the normal mode oscillations can be used to determine the elusive angular momentum parameter of the black hole. The general formalism developed by diskoseismologists for relativistic systems can be readily applied to the Newtonian case of cataclysmic variables (CVs).  Some of these systems (e.g., the dwarf nova SS Cygni) show rapid oscillations in the UV with periods of tens of seconds and high coherence. In this paper, we assess the possibility that these dwarf nova oscillations (DNOs) are diskoseismic modes. Besides its importance in investigating the physical origin of DNOs, the present work could help us to answer the following question. To what extent are the similarities in the oscillation phenomenology of CVs and X-ray binaries (XRBs) indicative of a common physical mechanism?
\end{abstract}


\section{Introduction}

During the outburst phase (timescale $\sim$ a few days)
some non-magnetic ($B\ll 10^6$ G) CVs exhibit fluctuations with periods $P \gtrsim$ 10 s and $Q \equiv 1/|dP/dt| \sim 10^3 - 10^7$. The phenomenology of these DNOs is very rich \citep{w04}. Also observed are lpDNOs (long period DNOs): $P \gtrsim 30$ s and $Q \sim 10^3 - 10^7$, as well as QPOs (quasi-periodic oscillations): $P \gtrsim 100$ s  and $Q \sim 10$. At times, all three features can be observed simultaneously.

The fact that $P \sim$ the Keplerian frequency at the radius of the white dwarf has motivated several candidate theoretical explanations for DNOs, ranging from `hot blobs' \citep{pop} to star oscillation modes \citep{pp} to explanations based on the dynamics of the boundary layer or some other type of transition region located where the disk touches the white dwarf. Examples of the last type include magnetic accretion onto a slipping belt at the equator \citep{ww02} and `spreading layer' models \citep{pb}. Hydrodynamic oscillations in and near the boundary layer have been investigated by \citet{ca85} and \citet{chv}. However, their analyses were local (plane-wave WKB)
in radius and height. Indeed, \citet{ca85} mention the need of a global analysis. 
None of these models is truly successful at the quantitative level, although magnetic models appear to be favored.

Lately, other proposals based on the dynamics of the accretion disk have appeared. For example, \citet{klaw} propose a nonlinear hydrodynamical disk resonance explanation for the oscillations. Part of the reason this approach is attractive is that it could explain the similarity of the phenomenology in CVs and XRBs.

Some DNOs exhibit a 1:2 (e.g., the period halving in SS Cyg) and in one case (VW Hyi) a 1:2:3 harmonic structure. In some sources, a 1:15 relation of the frequencies of a QPO and DNO is observed \citep{m02,ww05}, similar to that seen in XRBs. Some high-frequency QPOs in black hole XRBs are in a 3:2 ratio \citep{ak,mr06}. All the explanations listed above (except the hot blob and resonance models) are not applicable to black holes, given the absence of a surface (even though some could apply to neutron stars). Conversely, QPO explanations based on general relativistic effects are not applicable to the CV case. The only common explanation is based on hydrodynamical oscillations in the accretion disk (and/or corona).

With the intention of adding to this debate, this paper explores the possibility that the DNOs in CVs are caused by global hydrodynamical oscillations (normal modes) in the inner accretion disk. To this end we will use diskoseismology, a formalism previously developed to study small adiabatic perturbations in relativistic accretion disks [see, e.g., \citet{wa,wso}]. The DNOs would arise as the fluctuations modulate the outgoing radiation.
Although 
\cite{yamasaki} mention global hydrodynamical 
p--modes 
(trapped acoustic waves)
in accretion disks around white dwarfs,
these modes lie in a different region from the one studied in this paper;
in addition, the
physical analysis of these authors
is much simpler than ours, as they work with vertically integrated variables
and limit themselves to axisymmetric modes.

In section 2 we briefly discuss the unperturbed disk model and then outline diskoseismology in the general case. The rest of the paper applies diskoseismology to cataclysmic variables. An introduction (section 3) precedes the formal WKB solution in section 4. Finally, section 5 presents the results and discusses their implications.

\section{Diskoseismology}

We take $c=1$, and express all distances in units of $GM/c^2$ and all frequencies in units of $c^3/GM$ (where $M$ is the mass of the white dwarf) unless otherwise indicated.  We work in cylindrical coordinates $r, \varphi, z$.

The stationary ($\partial/\partial t=0$), symmetric about the midplane $z=0$, and axially symmetric ($\partial/\partial\varphi=0$) unperturbed disk is taken to be described by the standard thin disk models 
for the Newtonian \citep{ss} and the relativistic \citep{nt} cases.
In particular, we note that the velocity components $v^r=v^z=0$ (the disk fluid moves in nearly circular orbits), the disk semi-thickness $h(r)\sim c_s/\Omega\ll r$ (where $c_s$ and $\Omega$ are the speed of sound and the Keplerian frequency, respectively), the disk is optically thick and radiatively efficient,
the viscous stress is given by $\alpha p$ (i.e., the pressure multiplied by a constant parameter), and typical values for the disk density and central object mass make the neglect of self-gravity, at both the unperturbed and the perturbed level, a good approximation. Further properties of the unperturbed disk will be discussed in section 5.

Akin to helioseismology, diskoseismology is the theoretical study of the normal modes of small adiabatic hydrodynamical global perturbations in geometrically thin, optically thick accretion disks. In this section, we describe diskoseismology in general terms. In the following sections, we will use Newtonian 
diskoseismology to study oscillations in white dwarf accretion disks.
Fortunately, the relativistic and nonrelativistic models are
formally very similar.

Even though the unperturbed disk is not in equilibrium,
this perturbative approach is justified whenever
the viscous time scales
during which the disk evolves as a whole are much longer than the modes' periods,
and whenever
the turbulence has spatial scales which are smaller than the modes'
spatial scales,
and 
it is quasi-steady on the time scales of
the modes.
(These conditions hold for the modes dealt with in this paper.)

We consider barotropic disks [$p=p(\rho)$, giving vanishing buoyancy frequency] and assume that the results thus obtained are generic, since typically the buoyancy frequency is much less than the characteristic dynamical frequency $\Omega$. In this case hydrostatic equilibrium provides the vertical density and pressure profiles
\be
\rho=\rho_0(r)(1-y^2)^{g}\; ,\quad p=p_0(r)(1-y^2)^{g+1}\;, \quad g\equiv1/(\Gamma-1) >0 \; ,
\ee
where $\Gamma>1$ is the adiabatic index.  One has $\Gamma=4/3$ within any radiation pressure dominated region of the disk, and $\Gamma=5/3$ within any gas pressure dominated region (such as the white dwarf accretion disks). The disk surfaces are at $y=\pm1$, with $y$ related to the vertical coordinate $z$ by
\[ y={z\over h(r)}\,\sqrt{\Gamma-1\over2\Gamma} \; . \]

To investigate the eigenmodes of the disk oscillations starting from these assumptions, one applies the formalism that \citet{il91,il92} developed for perturbations of purely rotating perfect fluids,
starting from the equations of conservation of mass and momentum. One can thus express the Eulerian perturbations of all physical quantities through a single function $\delta V \propto \delta p/\rho$ 
(the Eulerian perturbation of the pressure divided by the density) which satisfies a second-order partial differential equation. Due to the stationary and axisymmetric background, the angular and time dependences are factored out as $\delta V = V(r,z)\exp[i(m\varphi + \sigma t)]$, where $\sigma$ is the eigenfrequency, and the governing PDE becomes
\begin{equation}  
\frac{\partial}{\partial r}\left(\frac{g^{rr}\rho}{\omega^2-\kappa^2}\frac{\partial V}{\partial r}\right) +
\frac{\partial}{\partial z}\left(\frac{\rho}{\omega^2}\frac{\partial V}{\partial z}\right) + 
\frac{\rho\beta^2}{c_s^2}\,V = 0 \; ,
\label{PDE}
\end{equation}
where $\beta = dt/d\tau$, $g^{rr}$ is the metric coefficient, $c_s$ stands for the speed of sound, the corotation frequency
\be\label{corotation}
\omega \equiv \sigma + m \Omega \; ,
\ee
$\Omega(r)$ is the Keplerian frequency, and $\kappa(r)$ is the radial epicyclic frequency
(i.e., the frequency of radial perturbations of free particle circular orbits).

An assumption of strong variation of modes in the radial direction (characteristic radial wavelength $\lambda_r \ll r$) ensures the approximate WKB separability of variables in this PDE, $V(r,z) = V_r(r)V_y(r,y)$. The `vertical' part, $V_y$, of the functional amplitude $V(r,z)$ varies slowly with $r$. 
They obey \citep[][hereafter referred to as RD1]{per}
\bea
(1-y^2)\,\frac{d^2V_y}{d y^2} - {2gy}\,\frac{d V_y}{d y} + {2g\omega_*^2}\,
\left[1 -\left(1-\frac{\Psi}{\omega_*^2}\right)\left(1-y^2\right)\right]V_y 
= 0 \; ,
\label{rde1} \\
\frac{d^2 V_r}{dr^2} - \frac{1}{(\omega^2-\kappa^2)} \left[\frac{d}{dr}
(\omega^2-\kappa^2)\right]\frac{dV_r}{dr} + 
\beta^2 g_{rr} c_s^{-2} (\omega^2-\kappa^2)\left(1 -
\frac{\Psi}{\omega_*^2}\right)V_r = 0 \; ,  \label{rde2}
\eea
where
\be
\omega_* \equiv \omega/\Omega_\perp \; ;  \label{omegastar}
\ee
$\Omega_\perp(r)$ refers to the vertical epicyclic frequency
(i.e., the frequency of vertical perturbations of free particle circular orbits),
and the speed of sound $c_s$ is evaluated at midplane.
In these equations $\Psi(r)$ is the slowly-varying separation function.

These equations have been used in the past to reveal the properties of different types of modes in relativistic accretion disks. In this paper we solve the equations in the Newtonian limit. We employ, as in the original papers, WKB methods in a straightforward manner.

\section{Diskoseismic Modes for the White Dwarf Case: Physical Parameters and Scope}

We now apply diskoseismology to the accretion disks of white dwarfs. We are thus working in the Newtonian limit (for a mass $M=M_\sun$ white dwarf, its radius $R \sim 5\times 10^3$, corresponding to $7.4\times 10^3$ km), 
which amounts to setting
$g^{rr} = \beta = 1$, $g = 3/2$ in the equations of the previous section.
We note in particular that
\be\label{freqs}
  \Omega(r) = \kappa(r) = \Omega_\perp(r) = 1/r^{3/2}  \, ,
\ee
i.e., the rotational, radial epicyclic, and vertical epicyclic frequencies are all equal (up to corrections of relative size $\sim 1/r$ or smaller). 
As we will see, this results in a qualitative change in the nature of the 
solutions with respect to the relativistic case.

This is a straightforward problem with few free parameters. In addition to the white dwarf mass $M$, there is $r_i\;[\ge R(M)]$, the inner radius of the accretion disk. Then, one needs a functional form for the speed of sound. It is fortunate that the properties of the unperturbed disk (viscosity and luminosity) enter into the formalism only through the speed of sound, and that the dependence of the speed of sound on these properties is rather weak.

In this exploratory calculation, we ignore the complications related to the nature of the transition from the disk to the star surface or magnetosphere and assume that DNO oscillations are confined to the disk. We introduce a disk inner boundary condition parameter, given our ignorance of the physical conditions in the (magnetic) boundary layer. As we will see, the results are rather insensitive to this parameter.
The approximate character of our analysis should be stressed.
A deeper study of boundary layer dynamics
is needed in order to understand better the physical conditions that 
would allow for the existence of the modes described in this paper.
For instance, one could ask
how significant is the damping and `leakage' of the modes into the magnetosphere, where the radial epicyclic frequency is given by the more general expression 
$\kappa^2 = 2\Omega r^{-1} d(r^2 \Omega)/dr$, if the modes are still confined to a small vertical extent $h \ll r$.

As in the case of relativistic diskoseismology (with black holes or weakly magnetic neutron stars), the
formalism reveals the existence of different types of modes. In this first diskoseismic investigation of cataclysmic variables, we choose to study the g--modes, since they are trapped near the inner edge of the disk and are therefore the most luminous (and robust). By definition, the g--modes have corotating eigenfrequencies which are smaller than the gravitational radial oscillation (epicyclic) frequencies at the involved radii. In other words, the (smaller) pressure restoring forces act against the gravitational restoring forces. (The `g' in the name comes from a mathematical similarity to the g--modes within slowly rotating stars.) The acoustic (p--) modes extend to the outer region of the disk, thus involving strong effects of viscosity and the uncertain boundary conditions at its outer radius.

\section{Formal WKB Solution of the Eigenvalue Problem for G--modes}

In this section we will proceed to solve formally the eigenvalue problem in 
order to obtain equation (\ref{master}), our `master' equation, from which all 
further results in the paper are derived.

Taking into account expressions (\ref{freqs}), the separated equations (\ref{rde1}) and (\ref{rde2})
reduce in the Newtonian limit to
\bea
(1-y^2)\,\frac{d^2V_y}{d y^2} - {3y}\,\frac{d V_y}{d y} + {3\omega_*^2}\,
\left[1 -\left(1-\frac{\Psi}{\omega_*^2}\right)\left(1-y^2\right)\right]V_y 
= 0 \; ,  \label{difeqy}
 \\
\frac{d^2 V_r}{dr^2} - \frac{1}{(\omega^2-\Omega^2)} \left[\frac{d}{dr}
(\omega^2-\Omega^2)\right]\frac{dV_r}{dr} +c_s^{-2}(\omega^2-\Omega^2)\left(1 -
\frac{\Psi}{\omega_*^2}\right)V_r = 0 \; . \label{difeqr}
\eea
(Recall from section 2 that $g = 3/2$ for white dwarf accretion disks.)
Equations (\ref{difeqy}) and (\ref{difeqr}) are two coupled eigenvalue equations, the eigenvalues being $\omega$ and $\Psi$ (or equivalently, $\sigma$ and $\Psi$). We will solve these equations within a straightforward WKB formalism as described by \citet{osw} (hereafter referred to as RD3). With respect to boundary conditions, it turns out that it is sufficient to require that the vertical eigenfunction and its derivative remain finite at the surface of the disk and to introduce an inner boundary parameter for the radial equation, as mentioned.

By definition, g--modes exist wherever $\omega^2 < \kappa^2 (=\Omega^2)$. This means that the mode exists between an inner radius (which is $r_i$, the inner radius of the accretion disk, 
for the cases studied in this paper)
and the radius $r_+$, defined as the radius at which $|\omega| = \Omega$. 
The mode decays exponentially for $r > r_+$.

\subsection{WKB Solution of the Radial Eigenvalue Problem}

We will cast equation (\ref{difeqr}) in a form amenable to a WKB solution,
using a procedure which is similar to the one developed in RD3. 
We introduce the dependent variable
\be
 W \equiv \frac{1}{\Omega^2 - \omega^2}\frac{dV_r}{dr} \; ,
\ee
and the independent variable
\be
 \tau \equiv \int_r^{r_+} c_s^{-2}(r^\prime)
\left[  \frac{\Psi(r^\prime)}{\omega_*^2(r^\prime)} - 1 \right] dr^\prime \; ,
\ee
so that equation (\ref{difeqr}) can be written in the simpler form
\be \label{Wequation}
 \frac{d^2W}{d\tau^2} + S(\tau) W = 0 \; , \quad
S(\tau) \equiv \frac{c_s^2 (\Omega^2 - \omega^2)}{\Psi/\omega_*^2 - 1} \; .
\ee
The mode exists in the range $r_i < r < r_+$
or, equivalently, $0 < \tau < \tau_i \equiv \tau(r_i)$.
The function $S(\tau)$ changes sign at $\tau = 0$ (where $r = r_+$).
The boundary conditions are that $W$ decays for $\tau < 0$,
while at the other extreme, at $\tau = \tau_i$,
we parameterize our ignorance at the inner boundary by means of the
quantity $\theta_i$:
\be
 (\Omega^2 - \omega^2) \cos \theta_i W - \sin \theta_i \frac{dW}{d\tau} = 0 
\; , \ee
which is equivalent to
\be\label{boundarycondition}
  \cos \theta_i \frac{dV_r}{dr} - \sin \theta_i V_r = 0 \; .
\ee
A straightforward WKB solution can be obtained for equation (\ref{Wequation})
at all points except for $\tau$ close to $0$: 
\be\label{sol1}
 W \propto S^{-1/4}(\tau) \cos
\left[ \int_\tau^{\tau_+} S^{1/2}(\tau^\prime) d\tau^\prime + \Phi
\right] \; ,
\ee
where $\Phi$ is a constant to be determined.
For $\tau$ near zero, the solution is given by
the Airy function of the first kind:
\be\label{sol2}
 W \propto Ai(-s_+^{1/3} \tau) \; , \quad
 s_+ \equiv dS(0)/d\tau > 0  \; .
\ee
The asymptotic
matching 
of solutions (\ref{sol1}) and (\ref{sol2}) 
in the usual manner yields $\Phi = \pi/4 + n\pi$, where
$n$ is an integer, and
\be\label{wkbrad}
\int^{r_+(\sigma)}_{r_i} c_s^{-1}(r) \sqrt{(\Omega^2 - \omega^2)
(\Psi/\omega_*^2 - 1)} dr = \pi (n \pm 1/4)
\ee
for the cases $\theta_i = \pi/2$ and $0$, respectively.
Thus $n$ is the radial mode number constrained by the fact
that the integral in equation (\ref{wkbrad}) is positive.

\subsection{WKB Solution of the Vertical Eigenvalue Problem}

Using the new independent variable
\be
 \zeta \equiv \int_0^y 
(1-y^{\prime 2})^{-3/2}
dy^\prime \; ,
\ee
we can write equation (\ref{difeqy}) in the simpler form
\be \label{Qeq}
 \frac{d^2V_y}{d\zeta^2} + Q(\zeta) V_y = 0 \; , \quad
Q(\zeta) \equiv 
{3\omega_*^2}\,
\left[1 + \left(\frac{\Psi}{\omega_*^2}-1\right)\left(1-y^2\right)\right]
\left(1-y^2\right)^2
\; ,
\ee 
which is amenable to a WKB solution.

We now discuss the issue of boundary conditions.
In view of the symmetry of the differential equation 
(\ref{difeqy}),
one has
\be
\frac{dV_y}{dy}\Biggl|_{y=0}=0 \qquad{\rm or}\qquad V_y\Biggl|_{y=0}=0
\;  \label{6.1}
\ee
for even and odd modes in the disk, respectively, and
it is enough to consider the interval $0<y<1$.
Near the singular boundary point $y=1$, 
equation (\ref{difeqy}) can be written 
\be 
\frac{2}{3}\frac{d^2V_y}{d y^2} 
- \left(\frac{1}{1-y}+\dots\right)\,\frac{d V_y}{d y} +
\left(\frac{\omega_*^2}{1-y}+\dots\right)\,V_y = 0 \; ,
\label{vertaty=1}
\ee
where the dots represent non-singular terms. 
According to the analytical theory of second order ODEs 
[see, e.g., \citet{olv}], the general solution to equation (\ref{difeqy}) 
near $y=1$ has the form
\be
V_y=C_1v_1(y)+C_2(1-y)^{-1/2}v_2(y) \; ,
\label{near1}
\ee
where $C_{1}$ and $C_2$ 
are arbitrary constants, and $v_{1}(y)$ and $v_2(y)$ 
are converging power series of $(1-y)$.
Imposing the boundary condition that the eigenfunction and its derivative
be finite at $y=1$ implies thus that $C_2$ = 0.
 
As noted in RD1 and implied by equation (\ref{vertaty=1}) 
and the boundary condition, 
the vertical eigenfunction near the boundary can be written in terms of the 
Bessel function of the first kind as
\be
V_y(y)\propto (1-y)^{-1/4}J_{1/2}(2|\omega_*|\sqrt{(3/2)(1-y)} \, )  \; ,
\label{6.4}
\ee
the asymptotic (large argument) expression of which is
\be
V_y(y)\propto (1-y)^{-1/2}\cos\left[|\omega_*|\sqrt{6(1-y)}-(\pi/2)\right] \; .\label{6.7}
\ee
On the other hand, equation (\ref{Qeq}) can be solved 
with a WKB method for 
values of $y$ not close to 1:
\be
 V_y \propto Q^{-1/4}(\tau) \cos
\left[ \int_0^{\zeta} Q^{1/2}(\zeta^\prime) d\zeta^\prime 
-{\cal I}\pi/2 
\right] \; ,
\ee
where ${\cal I}=0,1$ for the even and odd eigenfunctions, respectively.
Matching this with the asymptotic expression (\ref{6.7}) of the boundary solution (\ref{6.4}), we 
arrive at
\be\label{wkbver}
\int_0^1 \sqrt{\frac{1 + (\Psi/\omega_*^2 - 1)(1-y^2)}{1 - y^2}} dy
  = \frac{\pi J}{\sqrt{3} |\omega_*|}  \; .
\ee
The parity (of the vertical eigenfunction) and the vertical mode number $j$ 
are contained in a single quantity
$J \equiv j + 3/4 \pm 1/4$ for the odd and even cases, respectively. 
Thus $J$ takes positive integer and half-integer values (see Table 1).

It should be stressed that, because of the method used to solve the equations, the above results are accurate only when $\Psi/\omega_*^2 - 1 \gg 1$ 
for all $r$ in the range $r_i < r < r_+$.

\subsection{Uncoupling}

Equations (\ref{wkbrad}) and (\ref{wkbver}) are two eigenvalue equations
(the eigenvalues being $\Psi$ and $\omega$),
which need to be uncoupled. We start by inverting the integral $I$ on the left hand side of equation (\ref{wkbver}) to find $\Psi/\omega_*^2 - 1$ as a function of $I$:
\be
  \Psi/\omega_*^2 - 1 = f(I) \, .
\ee
Once this is done numerically, the function $f(I)$ can be substituted in equation (\ref{wkbrad}):
\be\label{master}
\int^{r_+}_{r_i} c_s^{-1}(r) \sqrt{(\Omega^2 - \omega^2)f(\pi J/ \sqrt{3}|\omega_*|)} dr = \pi (n \pm 1/4) \, .
\ee
The left--hand side of equation (\ref{master}) now only depends on the eigenfrequency $\sigma$ and not on $\Psi$ [recall definition (\ref{corotation})]. All further results in this paper are obtained from this equation.

\section{Results and Discussion}

In order to use equation (\ref{master}) and look for solutions with integer values of $n\ge 0$, we need a value for  $r_i$ and a functional form for the speed of sound $c_s$. We will take $r_i = 5000$, corresponding to  $7383 (M/M_\sun)$ km. The restriction $r_i\geq R$, the white dwarf radius, places a limit on its mass $M$. Using a standard mass--radius relation \citep{na}, this value of $r_i$ implies that $M\geq 0.76M_{\sun}$. 

As for the speed of sound, there are different solutions given by \citet{ss},
depending on the region of the disk in which one is working. 
For the values of interest in this paper, the relevant expression is 
the one corresponding to the outermost solution, 
in which temperatures are low enough that pressure is dominated by gas (not radiation) and opacity is dominated by photon absorption in the form of nonrelativistic thermal bremsstrahlung (`free-free' transitions).

Their solution for the speed of sound is given by
\be
(c_s/c)^{-1} = 300\, \alpha^{1/10} (M/M_\sun)^{1/4} \dot{M}_{16}^{-3/20}\, r^{3/8}  \; ,
\label{sound}
\ee
where $c$, $\alpha$, $M$, and $\dot{M}_{16}$ stand, respectively, for the speed of light, the (dimensionless) viscosity parameter, the mass of the white dwarf, and the mass accretion rate in units of $10^{16}$ g/s. The small powers of the three parameters render our results relatively insensitive to them. We remind the reader that the interior structure of the 
unperturbed accretion disk enters our formulation only through the speed of sound. As mentioned above, we parameterize our ignorance about the radial inner boundary by means of the quantity $\theta_i$ introduced in equation (\ref{boundarycondition}).

\subsection{Axisymmetric Modes}

The eigenmodes are described by the azimuthal, radial and vertical mode numbers $m$, $n$, $j$, plus the value of the parity of the vertical eigenfunction. We start by studying the $m = 0$ modes. For them, the condition $\Omega^2 - \omega^2 > 0$ becomes $\Omega^2 - \sigma^2 > 0$, which means that the modes are trapped between $r_i$ and $r_+ = |\sigma|^{-2/3}$.

Using the characteristic values $r_i = 5000$, $M = M_\sun$, $\alphav = 0.1$ \citep{lasota}, 
and $\dot{M}_{16} = 1$ \citep{fkr}, we calculate $\sigma$ for different values of $n$, $j$ and parity
for the case $\theta_i = \pi/2$.
(Note that the values of $j$ and parity are contained in $J$ in the manner described in section 4.2 and Table 1.) Table 2 shows the results. The dimensional frequency $f = 3.23\times 10^4 (M_\sun/M)|\sigma|$ Hz. The $n=0$, $J=1$ entry thus corresponds to a period of $12$ seconds.

Table 3 shows $\Delta r / r_i \equiv (r_+ - r_i)/r_i$ for the same values of the parameters. As in the case of axisymmetric g--modes in relativistic accretion disks (RD1), the value of $c_s/c$ is small enough that:  
a) the small $n$ eigenfrequencies are close to the maximum value of $\kappa(r)$ [which equals $\Omega(r_i) = 2.83\times 10^{-6}$ in the present case], approaching it as $j$ increases; b) the modes' radial width decreases as $j$ increases and increases as $n$ increases.

Self-consistency requires that the parameter $\Psi/\omega_*^2 - 1$ not be small, and the larger it is, the better the WKB approximation. It turns out that $\Psi/\omega_*^2 - 1 \sim 1$ when $J = 1$ and $\Psi/\omega_*^2 - 1 \gg 1$ when $J > 1$, so the procedure passes this test. (Incidentally, it is interesting that the minimum value of $\Psi/\omega_*^2 - 1$ over the allowed values of the radius does not depend on $n$ or $m$, only on $J$.)

There is no $J=1/2$ entry in Table 2 because there is no WKB solution. A direct numerical solution of equations
(\ref{difeqy}) and (\ref{difeqr}) shows that the absence of a solution is not an artifact of the WKB method, although it could result from the assumptions made in obtaining these equations. The possibility remains open that there are modes not detected by our methods.

Next we compute the behavior of the eigenfrequency $\sigma$ as the various parameters change. For the choice $J = 1$, $n = 0$, $\theta_i = \pi/2$, we obtain
\bea 
 \partial\log\sigma/\partial \log r_i &=& -1.5 \; , \\
 \partial\log\sigma/\partial\log\alphav &=& 0.0041 \; , \\
 \partial\log\sigma/\partial\log M &=& 0.010 \; , \label{sigM} \\
 \partial\log\sigma/\partial\log\dot{M} &=& -0.0061 \; , 
\eea
which means that the dependence of $\sigma$ on $\alpha$, $M$, and $\dot{M}$ can be neglected since it is much weaker than its (expected) dependence on $r_i$. Moreover, a change of $\theta_i$ from $\pi/2$ to $0$ implies only a change of a few percent in $\sigma$. Note, however, that there can be no $n = 0$ solution with $\theta_i = 0$ because then the RHS of equation (\ref{wkbrad}) would be negative.

Also note that equation (\ref{sigM}) assumes a fixed value of $r_i$. Care must be taken then when comparing with observations, since usually the parameters are not independent. For example, for a set of white dwarfs of fixed composition there is a relationship between their radius ($R$) and mass. So since $\sigma \cong \sigma(r_i)$ and  $r_i = r_i(M, \dot{M})$ (for fixed surface magnetic field), one has thus, for fixed $\dot{M}$, 
\be
\frac{\partial \log\sigma}{\partial \log M} = \frac{\partial \log r_i}{\partial \log M} 
\frac{d\log\sigma}{d\log r_i} = 0.50 \; ,
\ee
for the case $M \propto r_i^{-3}$ \citep{na} appropriate to low--mass white dwarfs with weak magnetic fields ($r_i=R$).

When computing the variation of $\sigma$ with the luminosity $L$ for any specific cataclysmic variable,
if one assumes that $r_i \propto L^{-\gamma}$, where $\gamma$ is a constant, one obtains
\be
 \frac{d\log P}{d \log L} = - \frac{\partial\log r_i}{\partial\log L}  
 \frac{d\log\sigma}{d\log r_i} = - 1.5 \; \gamma \; .
\ee
Such a dependence of $r_i$ on luminosity could be due to strong magnetic field effects \citep{st}. The idea is that inside the magnetosphere the accretion flow is controlled by magnetic fields, so one takes $r_i$ to be the equatorial radius of the magnetosphere. Its value should be only moderately greater than the white dwarf radius $R$ for those magnetic field strengths thought to be present in many of those white dwarf systems which exhibit DNOs. For example, the value of $d\log P/d\log L = -0.091$ for the dwarf nova SS Cygni reported by \citet{m96}  is reproduced by our formalism provided $\gamma = 0.06$.

In summary, the model may be consistent with observations. The radial width of the modes may be large enough to make them visible, and the periods are consistent, although this is hardly surprising as they are related to the Keplerian period at $r_i$. This leads to the limit $P\geq 10.94(M_{\sun}/M)^{1/2}[r_i(\mbox{km})/7383 \mbox{ km}]^{3/2}$ seconds on the periods of axisymmetric modes. However, the physics of the magnetosphere is at present too uncertain to provide a means to test the model accurately. At this point it makes sense to look at the higher $m$ modes.

\subsection{Non-Axisymmetric Modes}

Here we discuss the $m > 0$ modes. Note that from definition (\ref{corotation}) and the fact that only 
$|\omega|$ enters in the equations, we can, without loss of generality, consider only the non-negative $m$ cases provided we allow for negative $\sigma$.

The condition $\Omega^2 - \omega^2 > 0$ implies that
\be
-m - 1 < \sigma/\Omega(r) < - m + 1  \; .
\ee
Therefore, g--modes are trapped between $r_i$ and $r_+ = (m+1)^{2/3} |\sigma|^{-2/3}$. The inner boundary of the mode ($r_-$) is $r_i$, provided that $|\sigma| \geq (m - 1) \Omega(r).$
The negative sign of $\sigma$ reflects the fact that the oscillations' wavefronts travel in the same direction as the fluid.

Table 4 and Figure 1 show some properties of the $n = 0$, $J = 1$ modes for the $m = $ 0, 1 and 2 cases. Interestingly, the results are consistent with the observation of a 1:2:3 harmonic structure in the power spectra of the DNO VW Hyi, reported by \citet{ww05}. The observed peaks remain harmonic even during a threefold increase in the fundamental frequency. As in other models, the increase in frequency is due to a decrease in $r_i$ induced by an increase in $\dot{M}$ (and thus $L$).

If one expresses $\sigma$ in terms of $r_+$ and works to lowest order in $\Delta r/r_i$, from equation (\ref{master}) one obtains 
\be 
\Delta r/r_i \sim (n\pm 1/4)^{2/3}(m+1)^{-1/3}\epsilon^{2/3} \; .\label{delr}
\ee 
Since $\epsilon \equiv h(r_i)/r_i \sim (c_s/\Omega r)_{r_i}$, using equation (\ref{sound}) and our chosen set of parameters (which only weakly affect the estimate) gives $\epsilon\approx 4.2\times 10^{-3} \, r_i^{1/8} \approx 1.2\times 10^{-2}$. Then equation (\ref{delr}) is seen to agree approximately with the values in Tables 3 and 4.

The $m>0$ diskoseismic modes are in principle harder to observe than the axisymmetric modes, since in the former the modulation of the outgoing radiation is not as efficient due to cancelations over the entire disk (and their radial extent is somewhat smaller). However, the fact that the white dwarf can eclipse parts of the inner disk renders these modes more visible.
We predict that the energy spectrum of the DNOs is characteristic of the very inner disk (and not other regions such as the magnetosphere), and thus should be close to blackbody at the highest disk temperature.

As in the case of g--modes in relativistic accretion disks, the eigenfrequency splittings when $m$ changes by unity are much larger than the $n$ and $j$ splittings. The g--modes that are trapped by the general relativistic behavior of the radial epicyclic frequency $\kappa$ have $m$ splittings that are not as harmonic as those found here (Table 4). In particular, their eigenfrequencies are only proportional to $m$ for $m\gtrsim 2$ [an interesting difference between the relativistic and the nonrelativistic eigenfrequencies is that, for $m \gg 1$, $|\sigma| \rightarrow m \Omega(r_i)$ for the former (see RD1) while $|\sigma| \rightarrow (m+1) \Omega(r_i)$ for the latter].
However, it must be noted that it appears that g--modes with small mode numbers are viscously unstable, at least if the viscosity acts in the usual hydrodynamic way. 
This result is proven for the relativistic case in \citet{ort}, where the exponential growth rate of the mode is roughly $\alpha \Omega$. For the nonrelativistic g--modes studied in this paper, an application of  
their method yields the same result. Furthermore, these low--lying modes do not contain  a corotation resonance [where $\omega(r)=0$]. 

\section{Comments}

Thus, if the ideas in this paper are correct and DNOs are linear diskoseismic modes then the results could suggest a scenario in which white dwarf DNOs and black hole QPOs are produced through somewhat different physical mechanisms. After all, as \citet{klaw} themselves point out, the phenomenology is similar but not identical. While in white dwarf DNOs the simultaneous observation of two or even three of the harmonics is well established, in black holes usually only one mode of oscillation is observed at a given moment. Furthermore, white dwarf DNOs (and neutron star QPOs) have frequencies that vary with time, while QPOs in black holes have fixed frequencies. In neutron star XRBs, the (twin) high--frequency QPOs are analogous to the DNOs, but their frequency ratio is much closer to unity \citep{vdk}. It appears to be controlled (magnetically) by the neutron star, since their frequency difference is close to the neutron star spin frequency or half of it. We note that some DNOs are split at a QPO frequency. We also note that with increasing radius of a neutron star magnetosphere, their g-modes should approach (for $r_i\gtrsim 8$) those of this paper.

We hope to investigate the magnetic boundary layer in more detail in a future paper. How much will the boundary layer contribute to the damping (or growth) of the g--modes? Further work should try to explain why the value of $Q$ is so high. Naively, one would expect $Q \sim 1/\alpha$. (It is fortunate that most of our results are fairly insensitive to the value of $\alpha$.) The nonlinear growth and coupling of the modes must also be investigated.

\acknowledgments

This work was supported by grant 075-2002 (Incentivos) of Ministerio de Ciencia y Tecnolog\'{\i}a, Costa Rica, grant 829-A3-078 of Vicerrector\'{\i}a de Investigaci\'on, Universidad de Costa Rica, and NASA grant NAS 8-39225 to Gravity Probe B. We are also grateful to the Aspen Center for Physics for its support.

\clearpage

\begin{deluxetable}{cccccc}
\tablecolumns{3}
\tablewidth{0pc}
\tablecaption{Value of $J$ as a function of $j$ and parity}
\noindent
\tablehead{
\colhead{$j$} &
\colhead{parity} &
\colhead{$J$}}
\startdata
$0$ & even & $1/2$   \\
$0$ & odd &  $1$   \\
$1$ & even & $3/2$    \\
$1$ & odd & $2$       \\
\enddata
\end{deluxetable}

\begin{deluxetable}{cccccc}
\tablecolumns{5}
\tablewidth{0pc}
\tablecaption{Values of the eigenfrequency $\sigma \times 10^6$ for $m=0$ 
g--modes.}
\noindent
\tablehead{
\colhead{} &
\colhead{} &
\multicolumn{3}{c}{$J$} \\
\cline{3-5} \\
\colhead{$n$} &
\colhead{ }   &
\colhead{$1$}    &
\colhead{$3/2$}    &
\colhead{$2$}    
  }
\startdata
$0$ & & 2.65 & 2.73 & 2.75   \\
$1$ & & 2.39 & 2.55 & 2.61   \\
$2$ & & 2.23 & 2.43 & 2.51   \\
\enddata
\end{deluxetable}

\begin{deluxetable}{cccccc}
\tablecolumns{5}
\tablewidth{0pc}
\tablecaption{Values of the modes' fractional radial width
$\Delta r / r_i$ for $m=0$ 
g--modes}
\noindent
\tablehead{
\colhead{} &
\colhead{} &
\multicolumn{3}{c}{$J$} \\
\cline{3-5} \\
\colhead{$n$} &
\colhead{ }   &
\colhead{$1$}    &
\colhead{$3/2$}    &
\colhead{$2$}    
  }
\startdata
$0$ & & 0.044 & 0.024 & 0.019   \\
$1$ & & 0.12\phantom{0} & 0.072 & 0.055   \\
$2$ & & 0.17\phantom{0} & 0.11\phantom{0} & 0.083   \\
\enddata
\end{deluxetable}

\begin{deluxetable}{cccccc}
\tablecolumns{5}
\tablewidth{0pc}
\tablecaption{Harmonic structure of the
$J=1$, $n=0$ g--modes}
\noindent
\tablehead{
\colhead{$m$} &
\colhead{ }   &
\colhead{$-10^6 \,\sigma_m$}    &
\colhead{$\sigma_m/\sigma_0$}    &
\colhead{$\Delta r/r_i$}    
  }
\startdata
$0$ & & 2.65     & 1\phantom{.00} & 0.044   \\
$1$ & & 5.39     & 2.03           & 0.033   \\
$2$ & & 8.14     & 3.07           & 0.028   \\
\enddata
\end{deluxetable}

\clearpage

\begin{figure}
\figurenum{1}
\epsscale{1}
\plotone{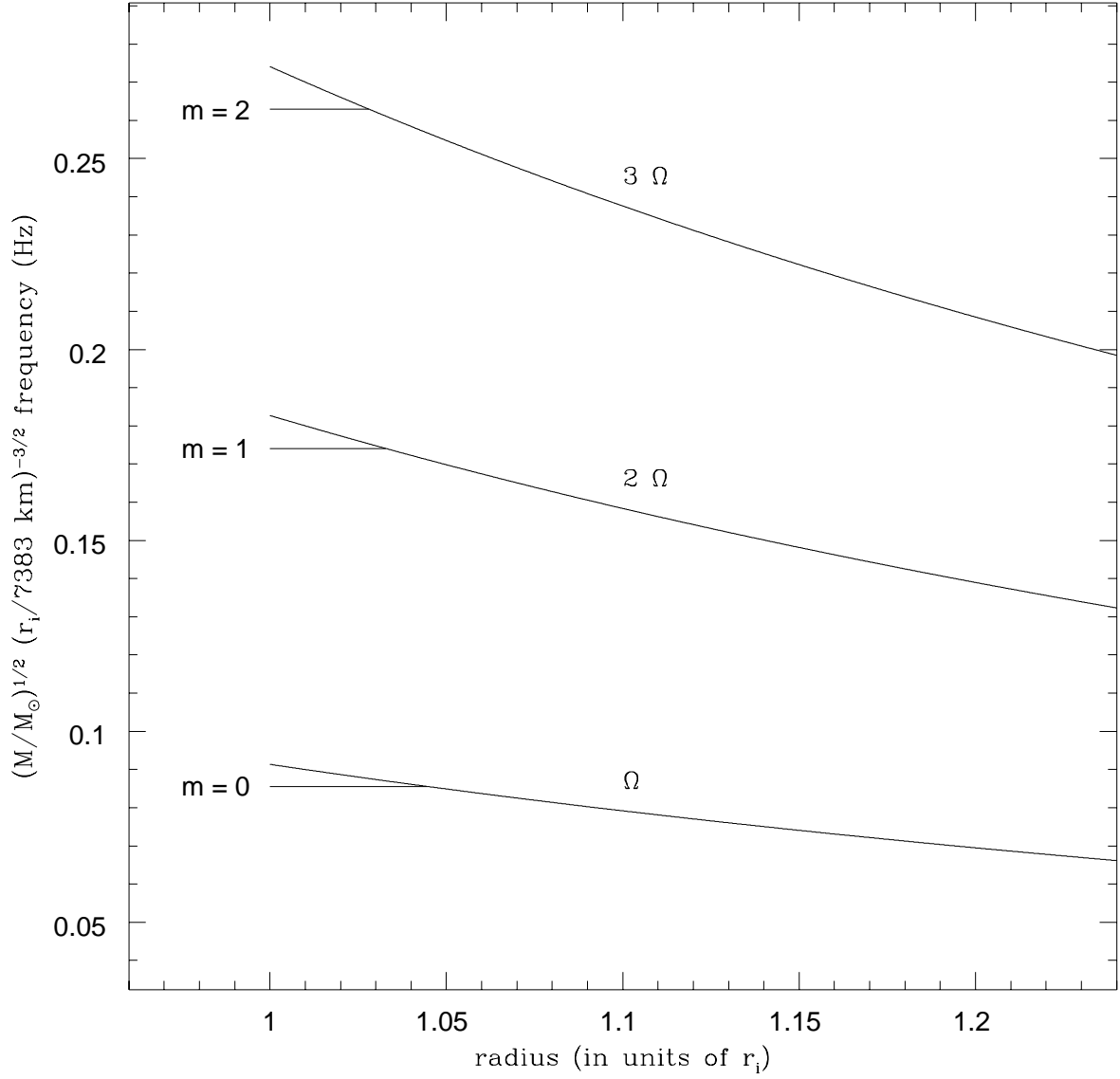}
\caption{The horizontal lines show the radial extent and the frequency of the three lowest fundamental ($n=0,J=1$) axial g-modes.}
\end{figure}

\end{document}